\documentclass[prd,preprint,superscriptaddress,showpacs,byrevtex]{revtex4}
\usepackage{epsfig}

\newcommand{\be}{\begin{equation}}
\newcommand{\ee}{\end{equation}}
\newcommand{\bea}{\begin{eqnarray}}
\newcommand{\eea}{\end{eqnarray}}

\begin{document}
%
%
\title{ Schwinger Mechanism for Quark-Antiquark Production
in the Presence of Arbitrary Time Dependent Chromo-Electric Field }
\author{Gouranga C. Nayak} \email{nayak@max2.physics.sunysb.edu}
\affiliation{C. N. Yang Institute for Theoretical Physics, Stony Brook
University, SUNY, Stony Brook, NY 11794-3840, USA }
%
%
\begin{abstract}

We study the Schwinger mechanism in QCD in the presence of
an arbitrary time-dependent chromo-electric background field
$E^a(t)$ with arbitrary color index $a$=1,2,...8 in SU(3).
We obtain an exact result for the non-perturbative
quark (antiquark) production from an arbitrary $E^a(t)$ by directly
evaluating the path integral. We find that the exact result
is independent of all the time derivatives $\frac{d^nE^a(t)}{dt^n}$
where $n=1,2,...\infty$. This result has the same functional dependence
on two Casimir invariants $[E^a(t)E^a(t)]$ and $[d_{abc}E^a(t)E^b(t)E^c(t)]^2$
as the constant chromo-electric field $E^a$ result with the
replacement: $E^a \rightarrow E^a(t)$. This result relies crucially on the
validity of the shift conjecture, which has not yet been established.

\end{abstract}
\maketitle

\newpage

In 1951 Schwinger derived an exact one-loop
non-perturbative result for electron-positron pair production in QED
from a constant electric field $E$ by using the proper time method
\cite{schw}. In QCD this result depends on two independent Casimir
invariants in SU(3): $C_1=[E^aE^a]$ and $C_2=[d_{abc}E^aE^bE^c]^2$
where $a,b,c$=1,2,...8 \cite{nayak1,nayak2}. Recently we have studied
the Schwinger mechanism for gluon pair production in the presence of
arbitrary time dependent chromo-electric field in \cite{gluon,gluon1}.
This technique is also applied in \cite{quark} to study path integration in QCD
in the presence of arbitrary space-dependent (one dimensional)
static color potential. This result relies crucially on the validity of
the shift conjecture \cite{nayak4}, which has not yet been established.

In this paper we study the Schwinger mechanism for quark-antiquark
production in the presence of an
arbitrary time-dependent chromo-electric background field
$E^a(t)$ with arbitrary color index $a$=1,2,...8 in SU(3).
We obtain an exact non-perturbative result for quark (antiquark)
production from arbitrary $E^a(t)$ by directly evaluating the path
integral.

We obtain the following exact non-perturbative
result for the probability of quark (antiquark)
production per unit time, per unit volume and per unit transverse momentum
from an arbitrary time dependent
chromo-electric field $E^a(t)$ with arbitrary color index $a$=1,2,...8 in SU(3):
\bea
\frac{dW_{q (\bar q)}}{dt d^3x d^2p_T}~
=~-\frac{1}{4\pi^3} ~~ \sum_{j=1}^3 ~
~|g\Lambda_j(t)|~{\rm ln}[1~-~e^{-\frac{ \pi (p_T^2+m^2)}{|g\Lambda_j(t)|}}].
\label{1}
\eea
In the above equation $m$ is the mass of the quark and
\bea
\Lambda_1(t)~=~\sqrt{\frac{C_1(t)}{3}}~{\rm cos}\theta(t);  ~~~~~
\Lambda_{2,3}(t)~=~\sqrt{\frac{C_1(t)}{3}}~{\rm cos}~ (2\pi/3 \pm \theta(t));~~~~~~
{\rm cos}^23\theta(t)=\frac{3C_2(t)}{C_1^3(t)},~~~
\label{lm}
\eea
where
\bea
C_1(t)=[E^a(t)E^a(t)];~~ C_2(t)=[d_{abc}E^a(t)E^b(t)E^c(t)]^2
\label{cas}
\eea
are two independent time-dependent Casimir/gauge invariants in SU(3).

This result has the remarkable feature that it is independent of all the time
derivatives $\frac{d^nE^a(t)}{dt^n}$ and
has the same functional form as the constant chromo-electric field $E^a$
result \cite{nayak1} with: $E^a \rightarrow E^a(t)$.

Now we will present a derivation of eq. (\ref{1}).

The Lagrangian density for a
quark (antiquark) in the presence of background chromo field $A_\mu^a(x)$ is given by
\bea
{\cal L} ={\bar \psi}^j(x)[\delta_{jk}{\hat p}\!\!\!\slash -gT^a_{jk}A\!\!\!\slash^a(x)-\delta_{jk}m]\psi^k(x)= {\bar \psi}^j(x)M_{jk}[A]\psi^k(x)
\eea
where $a$=1,2,...8 and $j,k$=1,2,3.
The vacuum-to-vacuum transition amplitude is given by
\bea
<0|0>=\frac{Z[A]}{Z[0]}= \frac{\int [d\bar{\psi}][d\psi] e^{i\int d^4x~{\bar \psi}^j(x)M_{jk}[A]\psi^k(x)}}{
\int [d\bar{\psi}][d\psi] e^{i\int d^4x~{\bar \psi}^j(x)M_{jk}[0] \psi^k(x)}} =
{\rm \frac{Det[M[A]]}{Det[M[0]]} } = e^{iS}
\label{gen}
\eea
which gives
\bea
S=-i{\rm Tr~ln}[\delta_{jk}{\hat p}\!\!\!\slash -gT^a_{jk}A\!\!\!\slash^a(x)-\delta_{jk}m]+i{\rm Tr~ln}[\delta_{jk}{\hat p}\!\!\!\slash -\delta_{jk}m].
\label{s1}
\eea
Since the trace is invariant under transposition we find
\bea
S=-i{\rm Tr~ln}[\delta_{jk}{\hat p}\!\!\!\slash -gT^a_{jk}A\!\!\!\slash^a(x)+\delta_{jk}m]+i{\rm Tr~ln}[\delta_{jk}{\hat p}\!\!\!\slash +\delta_{jk}m].
\label{s2}
\eea
Adding eqs. (\ref{s1}) and (\ref{s2}) we find
\bea
S=-\frac{i}{2} {\rm Tr}~
{\rm ln}[(\delta_{jk} \hat{{p}\!\!\!\slash} -gT^a_{jk}{{A}\!\!\!\slash}^a(x))^2-\delta_{jk}m^2] +\frac{i}{2}{\rm Tr~ ln}[\delta_{jk}(\hat{p}^2  -m^2)]
\label{dir}
\eea
where
\bea
{\rm Tr} {\cal O} =
{\rm tr}_{\rm Dirac} {\rm tr}_{\rm color}\int d^4x <x|{\cal O} |x>
\eea
Since it is convenient to work with the exponential of the trace we write
\bea
{\rm ln}(\frac{a}{b})= \int_0^\infty \frac{ds}{s}[ e^{-is(b-i \epsilon)}
-e^{-is(a-i \epsilon)}].
\eea
Hence we find from eq. (\ref{dir})
\bea
&& S=\frac{i}{2} {\rm tr}_{\rm Dirac} {\rm tr}_{\rm color}\int d^4x \nonumber \\
&&<x| \int_0^\infty \frac{ds}{s}
[e^{-is[(\delta_{jk} \hat{{p}} -gT^a_{jk}A^a(x))^2+ \frac{g}{2}\sigma^{\mu \nu}T^a_{jk}F^a_{ \mu \nu} -\delta_{jk}m^2-i\epsilon]} -e^{-is[\delta_{jk}(\hat{p}^2  -m^2)-i\epsilon]}]|x>.
\label{3c}
\eea
We assume the arbitrary time dependent chromo-electric field $E^a(t)$
to be along the beam direction (say along the z-axis)
and choose the axial gauge $A^a_3=0$ so that only
\bea
A^a_0(t,z) =-E^a(t)z
\label{7c}
\eea
is non-vanishing. Using eq. (\ref{7c}) in (\ref{3c}) and evaluating the Dirac trace by
using
\bea
(\gamma^0 \gamma^3)_{\rm eigenvalues} =(\lambda_1, \lambda_2, \lambda_3, \lambda_4) =(1,1,-1,-1)
\label{eigend}
\eea
we find
\bea
&& S =\frac{i}{2} \sum_{l=1}^4 {\rm tr}_{\rm color}~\int_0^\infty \frac{ds}{s}
\int dt <t| \int dx <x| \int dy <y| \int dz <z| \nonumber \\
&& e^{-is [(\frac{\delta_{jk}}{i}\frac{d}{dt} +gT^a_{jk}E^a(t) z)^2- \hat{p}_z^2- \hat{p}_T^2+i g\lambda_l T^a_{jk}E^a(t)-m^2-i\epsilon]} -e^{-is(\delta_{jk}(\hat{p}^2-m^2)-i\epsilon)} |z> |y> |x> |t>.~~
\eea
We write this in the color matrix notation
\bea
&& S =\frac{i}{2} \sum_{l=1}^4 {\rm tr}_{\rm color}~[\int_0^\infty \frac{ds}{s}
\int dt <t| \int dx <x| \int dy <y| \int dz <z| \nonumber \\
&& e^{-is [(\frac{1}{i}\frac{d}{dt} +gM(t) z)^2- \hat{p}_z^2- \hat{p}_T^2+i g\lambda_l M(t)-m^2-i\epsilon]} -e^{-is((\hat{p}^2-m^2)-i\epsilon)} |z> |y> |x> |t>]_{jk}
\eea
\bea
\noindent {\rm where}~~~~~~~~~~~~~~~~~~~~~~~M_{jk}(t)=T^a_{jk}E^a(t).~~~~~~~~~~~~~~~~~~~~~~~~~~~~~~~~~~~~~~~~~~~~~~~~~~~~~~~~~
\label{mt}
\eea
Inserting complete set of $|p_T>$ states (using $\int d^2p_T~ |p_T><p_T|=1$) we find from the
above equation
\bea
&& S^{(1)}=\frac{i}{2(2\pi)^2}  \sum_{l=1}^4
{\rm tr_{color}}[
\int_0^\infty\frac{ds}{s} \int d^2x_T\int d^2p_T
e^{is(p_T^2+m^2+i\epsilon)} \nonumber \\
&&~[ \int_{-\infty}^{+\infty} dt <t| \int_{-\infty}^{+\infty} dz <z|
 e^{-is[(\frac{1}{i} \frac{d}{dt} +gM(t) z)^2-\hat{p}_z^2 +ig\lambda_l M(t)]}
|z>|t> - \int dt  \int dz  \frac{1}{4\pi s}]
]_{jk}
\nonumber \\
\label{2ja}
\eea
where we have used the normalization $<q|p>=\frac{1}{\sqrt{2\pi}} e^{iqp}$.
At this stage we use the shift theorem \cite{nayak4} and find
\bea
&& S^{(1)}=\frac{i}{2(2\pi)^2}  \sum_{l=1}^4
{\rm tr_{color}}[
\int_0^\infty\frac{ds}{s} \int d^2x_T\int d^2p_T
e^{is(p_T^2+m^2+i\epsilon)} [ \int_{-\infty}^{+\infty} dt <t| \int_{-\infty}^{+\infty} dz \nonumber \\
&& <z+\frac{i}{gM(t)}\frac{d}{dt}|
 e^{-is[g^2M^2(t) z^2-\hat{p}_z^2 +ig\lambda_l M(t)]} |z+\frac{i}{gM(t)}\frac{d}{dt}>|t> - \int dt  \int dz  \frac{1}{4\pi s}]
]_{jk}
\nonumber \\
\label{12a}
\eea
where the $z$ integration must be performed from $-\infty$ to $+\infty$ for the shift theorem to be applicable.

Note that a state vector $|z+\frac{i}{a(t)}\frac{d}{dt}>$ which contains
a derivative operator is not familiar in physics.
However, the state vector $|z+\frac{i}{a(t)}\frac{d}{dt}>$
contains the derivative $\frac{d}{dt}$ not $\frac{d}{dz}$. Hence the
state vector is defined in the $z$-space with $\frac{d}{dt}$ acting as a c-number shift
in $z$-coordinate (not a $c$-number shift in $t$-coordinate). To see how one operates with such state
vector we find
\bea
<z+\frac{i}{a(t)}\frac{d}{dt}| p_z> f(t) =
\frac{1}{\sqrt{2\pi}} e^{i(z+\frac{i}{a(t)}\frac{d}{dt}) p_z} f(t) =
\frac{1}{\sqrt{2\pi}} e^{izp_z}e^{-\frac{p_z}{a(t)}\frac{d}{dt}} f(t).
\eea

Inserting complete sets of $|p_z>$ states (using $\int dp_z ~|p_z><p_z|=1$) in eq. (\ref{12a}) we find
\bea
&& S^{(1)}
=\frac{i}{2(2\pi)^2} \sum_{l=1}^4
\int_0^\infty \frac{ds}{s} \int d^2x_T\int d^2p_T
e^{is(p_T^2+m^2+i\epsilon)} [ F_l(s) - \int dt \int dz~ \frac{3}{4 \pi s}]
\label{12}
\eea
where
\bea
&& F_l(s)=\frac{1}{(2\pi)}
{\rm tr_{color}}~[
\int_{-\infty}^{+\infty} dt <t|
\int_{-\infty}^{+\infty} dz \int dp_z \int dp'_z~ e^{izp_z}e^{-\frac{1}{gM(t)}\frac{d}{dt}p_z} \nonumber \\
&& <p_z| e^{is[-g^2M^2(t) z^2+\hat{p}_z^2-i\lambda_l gM(t)]}|p'_z>
e^{\frac{1}{gM(t)}\frac{d}{dt} p'_z} e^{-izp'_z}|t> ]_{jk}.
\label{gn}
\eea
It can be seen that the exponential $e^{-\frac{1}{gM(t)}\frac{d}{dt}p_z}$
contains the derivative $\frac{d}{dt}$ which operates on
$<p_z|e^{is[-g^2M^2(t) z^2+\hat{p}_z^2-i\lambda_l gM(t)]}|p'_z>$
hence we can not move $e^{-\frac{1}{gM(t)}\frac{d}{dt}p_z}$ to the right.
We insert more complete set of $t$ states to find
\bea
&& F_l(s)=\frac{1}{(2\pi)}
{\rm tr_{color}}~[
\int_{-\infty}^{+\infty} dt
\int_{-\infty}^{+\infty} dz \int dp_z \int dp'_z~ \int dt' \int dt''
<t|e^{izp_z}e^{-\frac{1}{gM(t)}\frac{d}{dt}p_z}|t'> \nonumber \\
&& <t'|<p_z| e^{is[-g^2M^2(t) z^2+\hat{p}_z^2-i\lambda_l gM(t)]}|p'_z>  |t''><t''|e^{\frac{1}{gM(t)}\frac{d}{dt} p'_z} e^{-izp'_z}|t> ]_{jk}
\nonumber \\
&& =\frac{1}{(2\pi)} {\rm tr_{color}}~[ \int_{-\infty}^{+\infty} dt
\int_{-\infty}^{+\infty} dz \int dp_z \int dp'_z~ \int dt'
<t|e^{izp_z}e^{-\frac{1}{gM(t)}\frac{d}{dt}p_z}|t'> \nonumber \\
&& <p_z| e^{is[-g^2M^2(t') z^2+\hat{p}_z^2-i\lambda_l gM(t')]}|p'_z><t'|e^{\frac{1}{gM(t)}\frac{d}{dt} p'_z} e^{-izp'_z}|t> ]_{jk}.
\label{gnp}
\eea
Inserting more complete sets of states as appropriate we find
\bea
&& F_l(s) =\frac{1}{(2\pi)}
{\rm tr_{color}}~[
\int_{-\infty}^{+\infty} dt \int_{-\infty}^{+\infty} dz \int dp_z\int dp'_z \int dt' \int dp_0 \int dp'_0 \int dz' \int dz''  \nonumber \\
&& \int dp''_0 \int dp'''_0
<t|p_0> e^{izp_z}<p_0|e^{-\frac{1}{gM(t)}\frac{d}{dt}p_z}|p'_0><p'_0|t'>
<p_z|z'><z'| \nonumber \\
&&e^{is[-g^2M^2(t') z^2+\hat{p}_z^2-i\lambda_l gM(t')]}|z''><z''|p'_z>
<t'|p''_0><p''_0|e^{\frac{1}{gM(t)}\frac{d}{dt} p'_z} |p'''_0> e^{-izp'_z}<p'''_0|t> ]_{jk} \nonumber \\
&& =\frac{1}{(2\pi)^4} {\rm tr_{color}}~[
\int_{-\infty}^{+\infty} dt \int_{-\infty}^{+\infty} dz \int dp_z\int dp'_z \int dt' \int dp_0 \int dp'_0 \int dz' \int dz''  \nonumber \\
&& \int dp''_0 \int dp'''_0~
e^{itp_0} e^{izp_z}<p_0|e^{-\frac{1}{gM(t)}\frac{d}{dt}p_z}|p'_0>
e^{-it'p'_0} e^{-iz'p_z} \nonumber \\
&& <z'|e^{is[-g^2M^2(t') z^2+\hat{p}_z^2-i\lambda_l gM(t')]}|z''>
e^{iz''p'_z}
e^{it'p''_0}<p''_0|e^{\frac{1}{gM(t)}\frac{d}{dt} p'_z} |p'''_0> e^{-izp'_z}e^{-itp'''_0} ]_{jk}.
\label{gn151}
\eea
It can be seen that all the expressions in the above equation are independent
of $t$ except $e^{it(p_0-p'''_0)}$. This can be seen as follows
\bea
&& <p_0|f(t)\frac{d}{dt}|p'_0>= \int dt' \int dt'' \int dp''''_0 <p_0|t'> <t'|f(t)|t''><t''|p''''_0><p''''_0|\frac{d}{dt}|p'_0> \nonumber \\
&&= \int dt' \int dt'' \int dp''''_0 e^{-it' p_0}~ \delta(t'-t'') f(t'')
e^{it'' p''''_0}~
ip'_0~ \delta(p''''_0-p'_0)=ip'_0 \int dt'  ~e^{-it'(p_0-p'_0)}f(t') \nonumber \\
\label{indt}
\eea
which is independent of $t$ and $\frac{d}{dt}$.
Hence by using the cyclic property of trace we can take the matrix
$[<p''_0|e^{\frac{1}{gM(t)}\frac{d}{dt} p_z} |p'''_0>]_{jk}$ to the left.
The $t$ integration is now easy
($\int_{-\infty}^{+\infty} dt e^{it(p_0-p'''_0)}=2\pi \delta(p_0-p'''_0)$)
which gives
\bea
&& F_l(s) =\frac{1}{(2\pi)^3}
{\rm tr_{color}}~[  \int_{-\infty}^{+\infty} dz \int dp_z\int dp'_z
\int dt' \int dp_0 \int dp'_0  \int dz' \int dz''  \int dp''_0~
e^{izp_z} \nonumber \\
&& <p''_0|e^{\frac{1}{gM(t)}\frac{d}{dt} p'_z} |p_0><p_0|e^{-\frac{1}{gM(t)}\frac{d}{dt}p_z}|p'_0>e^{-iz' p_z} e^{-it'p'_0}
<z'| e^{is[-g^2M^2(t') z^2+\hat{p}_z^2-i\lambda_l gM(t')]}|z''> \nonumber \\
&&e^{it' p''_0} e^{iz''p'_z} e^{-izp'_z}
]_{jk}.
\label{gn51}
\eea
As advertised earlier we must integrate over $z$ from $-\infty$ to $+\infty$
for the shift theorem to be applicable \cite{nayak4}. The matrix element
$<z'| e^{-is[-g^2M^2(t) z^2+\hat{p}_z^2-i\lambda_l gM(t)]}|z''>$
is independent of the $z$ variable (it depends on $z'$ and $z''$ variables).
Hence we can perform the $z$ integration easily by using
$\int_{-\infty}^{+\infty} dz e^{iz(p_z-p'_z)} = 2\pi \delta(p_z-p'_z)$ to find
\bea
&& F_l(s) =\frac{1}{(2\pi)^2}
{\rm tr_{color}}~[ \int dp_z \int dt' \int dp_0 \int dp'_0 \int dz'
\int dz''  \int dp''_0 \nonumber \\
&& <p''_0|e^{\frac{1}{gM(t)}\frac{d}{dt} p_z} |p_0><p_0e^{-\frac{1}{gM(t)}\frac{d}{dt}p_z}|p'_0>e^{-iz' p_z}
e^{-ip'_0t'} <z'| e^{is[-g^2M^2(t') z^2+\hat{p}_z^2-i\lambda_l gM(t')]}|z''> \nonumber \\
&& e^{ip''_0t'}e^{iz''p_z}
]_{jk}.
\label{gn51b}
\eea
Using the completeness relation $\int dp_0 |p_0><p_0|=1$ we obtain
\bea
&& F_l(s) =\frac{1}{(2\pi)^2}
{\rm tr_{color}}~[\int dp_z \int dt' \int dp'_0 \int dz'  \int dz''  \nonumber \\
&& e^{-iz' p_z} <z'| e^{is[-g^2M^2(t') z^2+\hat{p}_z^2-i\lambda_l gM(t')]}|z''>  e^{iz''p_z}
]_{jk}.
\label{gn51f}
\eea
Since the color matrix $M_{jk}(t)=T^a_{jk}E^a(t)$ is antisymmetric it can
be diagonalized \cite{nayak1} by an orthogonal matrix $U_{jk}(t)$. The eigenvalues
\bea
[M_{jk}(t)]_{\rm eigenvalues}=[T^a_{jk}E^a(t)]_{\rm eigenvalues} =(\Lambda_1(t), ~\Lambda_2(t), ~\Lambda_3(t))
\label{eigenc}
\eea
can be found by evaluating the traces of $M(t)$, $M^2(t)$ and
${\rm Det} [M(t)$] respectively:
\bea
&& \Lambda_1(t)+\Lambda_2(t)+\Lambda_3(t)=0; ~~~~~~~~
\Lambda^2_1(t)+\Lambda^2_2(t)+\Lambda^2_3(t)=\frac{E^a(t)E^a(t)}{2}, \nonumber \\
&& \Lambda_1(t)\Lambda_2(t)\Lambda_3(t) =\frac{1}{12} [d_{abc}E^a(t)E^b(t)E^c(t)]
\label{tr1}
\eea
the solution of which is given by eq. (\ref{lm}).

Using these eigen values we perform the color trace in eq. (\ref{gn51f}) to find
\bea
F_l(s)=\frac{1}{(2\pi)^2} \sum_{j=1}^3 \int dp_z \int dt' \int dp'_0
\int dz'  \int dz''
e^{-iz'p_z} <z'| e^{is[-g^2\Lambda_j^2(t') z^2+\hat{p}_z^2-
i\lambda_l g\Lambda_j(t')]}|z''>
e^{iz'' p_z}.  \nonumber \\
\label{gn51i}
\eea
The above equation boils down to usual
harmonic oscillator, $\omega^2(t) z^2+\hat{p}_z^2$, with the
constant frequency $\omega$ replaced
by time dependent frequency $\omega(t)$.
The harmonic oscillator wave function
\bea
<z|n_t>=\psi_n(z)=(\frac{\omega(t)}{\pi})^{1/4}\frac{1}{(2^nn!)^{1/2}}
H_n(z \sqrt{\omega(t)})e^{-\frac{\omega(t)}{2}z^2}
\label{hmx}
\eea
(${\rm H}_n$ being the Hermite polynomial) is normalized
\bea
\int dz |<z|n_t>|^2=1.
\label{norm}
\eea
Inserting a complete set of harmonic oscillator states
(by using $\sum_n |n_t><n_t|=1$) in eq. (\ref{gn51i}) we find
\bea
 && F_l(s)=\frac{1}{(2\pi)^2} \sum_n \sum_{j=1}^3
  \int dp_z \int dt' \int dp'_0 \int dz'  \int dz''
 e^{-iz'p_z}
<z'|n_{t'}> e^{(-sg\Lambda_j(t')(2n+1)+s\lambda_l g\Lambda_j(t'))} \nonumber \\
&& <n_{t'}|z''> e^{iz'' p_z}
=\frac{1}{(2\pi)} \sum_n \sum_{j=1}^3 \int dt \int dp_0  \int dz~
|<z|n_t>|^2 e^{(-sg\Lambda_j(t)(2n+1)+s\lambda_l g\Lambda_j(t))} \nonumber \\
&& =\frac{1}{(2\pi)} \sum_n {\rm tr_{color}}~[
\int dt \int dp_0 ~e^{(-s(2n+1)gM(t) +sg\lambda_l M(t))}]_{jk}
\label{gn51l}
\eea
where we have used eq. (\ref{norm}). The Lorentz force equation in color space,
$\delta_{jk} dp_\mu = gT^a_{jk}F^a_{\mu \nu}dx^\nu$,
gives (when the chromo-electric field is along the $z$-axis, eq. (\ref{7c})),
$\delta_{jk} dp_0 = gT^a_{jk}E^a(t) dz = gM_{jk}(t) dz$. Using this in eq. (\ref{gn51l}) we obtain
\bea
&& F_l(s)=\frac{1}{(2\pi)} \sum_{j=1}^3
\int dt \int dz ~g \Lambda_j(t)
\frac{e^{sg\lambda_l \Lambda_j(t)}}{2 {\rm sinh(sg\Lambda_j(t))}}.
\label{gn51m}
\eea
Using this expression of $F_l(s)$ in eq. (\ref{12}) and summing over $l$ (by using the
eigen values of the Dirac matrix from eq. (\ref{eigend})) we find
\bea
S= \frac{i}{8 \pi^3} \sum_{j=1}^3
\int_0^\infty \frac{ds}{s} \int d^4x \int d^2p_T
e^{is(p_T^2+m^2+i\epsilon)}
[g\Lambda_j(t) ~\frac{{\rm cosh}(sg\Lambda_j(t))}{{\rm sinh}(sg\Lambda_j(t))}
-\frac{1}{s}].
\label{12ff}
\eea
The imaginary part of the above effective action gives real particle pair
production. The s-contour integration is straight forward \cite{schw,nayak1,nayak2,itzy}.
Using the series expansion
\bea
\frac{1}{{\rm sinh} x}=\frac{1}{x} + 2x \sum_{n=1}^\infty \frac{(-1)^n}{\pi^2 n^2 +x^2}
\eea
we perform the s-contour integration around the pole $s=\frac{in\pi}{|g\Lambda_j(t)|}$ to find
\bea
W=2 {\rm Im} S=
\frac{1}{4\pi^3}  \sum_{j=1}^3
\sum_{n=1}^\infty \frac{1}{n} \int d^4x \int d^2p_T |g\Lambda_j(t)|
e^{-n\pi \frac{(p_T^2+m^2)}{|g\Lambda_j(t)|}}.
\label{14ff}
\eea
Hence the probability of non-perturbative quark (antiquark)
production per unit time, per unit volume and per unit
transverse momentum from an arbitrary time dependent chromo-electric
field $E^a(t)$ with arbitrary color index $a$=1,2,...8 in SU(3) is given by
\bea
\frac{dW}{dt d^3x d^2p_T}~
=~-\frac{1}{4\pi^3} ~~ \sum_{j=1}^3 ~
~|g\Lambda_j(t)|~{\rm ln}[1~-~e^{-\frac{ \pi (p_T^2+m^2)}{|g\Lambda_j(t)|}}],
\label{1new}
\eea
which reproduces eq. (\ref{1}). The expressions for gauge invariant
$\Lambda_j(t)$'s are given in eq. (\ref{lm}).

To conclude we have obtained an exact non-perturbative result for quark-antiquark production
from arbitrary time-dependent chromo-electric field $E^a(t)$ with arbitrary
color index $a$=1,2,...8 in SU(3) via the Schwinger mechanism by directly
evaluating the path integral. This result relies crucially on the
validity of the shift conjecture, which has not yet been established.
We have found that the exact non-perturbative result
is independent of all the time derivatives $\frac{d^nE^a(t)}{dt^n}$
where $n=1,2,...\infty$ and has the same functional dependence on two
casimir invariants $[E^a(t)E^a(t)]$ and $[d_{abc}E^a(t)E^b(t)E^c(t)]^2$
as the constant chromo-electric field $E^a$ result \cite{nayak1} with the
replacement: $E^a \rightarrow E^a(t)$.

\acknowledgments

I thank Jack Smith and George Sterman for
careful reading of the manuscript. This work was supported in part by the
National Science Foundation, grants PHY-0354776 and PHY-0345822.

\end{document}